\documentclass{article}

\usepackage{arxiv}

\usepackage[utf8]{inputenc} 
\usepackage[T1]{fontenc}    
\usepackage{hyperref}       
\usepackage{url}            
\usepackage{booktabs}       
\usepackage{amsmath,amssymb,amsfonts}     
\usepackage{nicefrac}       
\usepackage{microtype}      
\usepackage{lipsum}
\usepackage{graphics}
\usepackage{graphicx}
\usepackage{mathrsfs}
\usepackage{multirow}
\usepackage{enumerate}
\usepackage{hyperref}
\usepackage{cleveref}
\usepackage{algorithm}
\usepackage{algorithmic}

\title{Approximate Uncertain Program}

\author{
  Xun Shen \\
  Department of Statistical Sciences\\
  The Graduate University for Advanced Studies\\
  Tokyo, Japan 106-8569\\
  \texttt{shenxun@ism.ac.jp} \\
   \And
 Jiancang Zhuang \\
  Institute of Statistical Mathematics\\
  Research Organization of Information and Systems\\
  Tokyo, Japan 190-8562\\
  \texttt{zhuangjc@ism.ac.jp} \\
  \And
  Xingguo Zhang \\
  Department of Mechanical Systems Engineering\\
  Tokyo University of Agriculture and Technology\\
  Tokyo, Japan, 184-8588\\
  \texttt{xgzhang@go.tuat.ac.jp} \\
}

\begin{document}
\maketitle

\begin{abstract}
Chance constrained program where one seeks to minimize an objective over decisions which satisfy randomly disturbed constraints with a given probability is computationally intractable. This paper proposes an approximate approach to address chance constrained program. Firstly, a single layer neural-network is used to approximate the function from decision domain to violation probability domain. The algorithm for updating parameters in single layer neural-network adopts sequential extreme learning machine. Based on the neural violation probability approximate model, a randomized algorithm is then proposed to approach the optimizer in the probabilistic feasible domain of decision. In the randomized algorithm, samples are extracted from decision domain uniformly at first. Then, violation probabilities of all samples are calculated according to neural violation probability approximate model. The ones with violation probability higher than the required level are discarded. The minimizer in the remained feasible decision samples is used to update sampling policy. The policy converges to the optimal feasible decision. Numerical simulations are implemented to validate the proposed method for non-convex problems comparing with scenario approach and parallel randomized algorithm. The result shows that proposed method have improved performance. 
\end{abstract}

\keywords{Chance constrained program, extreme learning machine, randomized optimization.}

\section{Introduction}
\label{sec:introduction}

Uncertain programs are optimization problems involved uncertainties in constraints or objectives\cite{Ben-Tal}. Uncertain program involves chance constraints or probabilistic constraints is identified as chance constrained program\cite{Prekopa}. Chance constraints are constraints within uncertain parameters which are required to hold with specified probability levles\cite{Charnes}. In recent 30 years, chance constrained program has been applied from economics and management fileds to various problems, such as model predictive control\cite{Schildbach}, automotive control\cite{Moser,Shen} , machine learning\cite{CampiClassification}, system identification and prediction\cite{CampiBook}.

However, chance constrained program is NP-hard due to the chance constraints. The challenge of solving chance constrained program directly motivates the development of approximation approach to solve it. In recent 20 years, the main stream has converged to scenario approach\cite{Calafiore,Luedtke} in which deterministic constraints imposed for finte sets of independently extracted samples of uncertain parameters are used to replace the chance constraints. Scenario approach preserves that the solution of the newly formulated deterministic program satisfies chance constraints with a determined bounds of probability\cite{Calafiore}. Afterwards, scenario approach with tight confidence bounds has been developed, in which a set of sampled constraints with tight confidence on violation probability is determined for approximating chance constraints. For instance, a certain proportion of parameter samples to define a set of sampled constraints and discard the rest can be used to approximate chance constraints with tight violation probabilities for fixed sample number \cite{CampiSampling}. A repetitive scenario approach that utilizes both priori and posteriori knowledge of probabilities can provide tighter bounds than approaches without prior information\cite{Cannon}. However, scenario approach still has fatal drawbacks. It cannot ensure that the obtained solution is the optimal one in the probabilistic feasible domain. when the sample number becomes larger, the obtained solution becomes more conservative and finally converges to the totally robust solution which is feasible for all uncertainty realizations. Moreover, the solution depends highly on the chosen samples of uncertain parameters, which might be guided to the wrong directions due to the bad choices of samples. 

Sample average approach has been proposed in \cite{Pagnoncelli} as a development of scenario approach. A sample average program is used as approximation of the chance constrained program in which the chance constraints are replaced by a meaure to indicates the violation probability. \cite{Geletu} extends sample average approach to an inner-outer approximate approach. However, sample average approach still needs too many assumptions on the program such as convexity of the function from decision to violation probability.

Bayesian optimization framework has been applied to optimization under unknown constraints recently\cite{Gramacy,Picheny}, which is essentially a data-driven approach for approximating the optimizer of the program. Statistical approach based on Gaussian processes and Bayesian learning to both approximate the unknown function and estimate the probability of meeting the constraints is developed to approximate the optimizer in unknown feasible domain. While, this can only be applied to expected constraints. Also, Gaussian process model is still not precise enough for approximating the feasible domain described by chance constraints.  

Stimulated by \cite{Pagnoncelli,Gramacy}, \cite{Shen2019} proposed a parallel randomized algorithm for chance constrained program in which randomized optimization and sample-discard method is applied to search the probabilistic feasible optimizer. As a development of \cite{Shen2019}, this paper addresses chance constrained program with a single neural networks-based approximate framework. The map from decision domain to violation probability domain is approximately modelled by Single Layer Feedforward Neural-networks (SLFNs). Based on the approximate violation probability map, randomized optimization algorithm is used to approach the optimizer in the probabilistic feasible domain of decision. The algorithm for updating parameters in single layer neural-network adopts sequential Extreme Learning Machine (ELM). Firstly, samples are extracted from both decision domain and random disturbance space. According to the chosen samples, violation probability map is updated. Then, in the second phase, the feasible decision samples are extracted uniformly but discarded with a portion according to the violation probability map. The remained feasible decision samples are used to calculated the corresponding objective values. These pairs are used to update sampling policy of the second layer. The policy converges to the optimal feasible decision.

The rest of the paper is organised as follows. Section \ref{sec:background} gives brief background of chance constrained program formerly and then present the problem description. The preliminaries for understanding the proposed method is given in Section \ref{sec:Preliminaries}. In Section \ref{sec:proposed}, sampling algorithms for violation probability map approximation and optimizer exploration is introduced. The numerical simulation for validating the sampling algorithm is presented in Section \ref{sec:numerical simulation}, using a non-convex program with chance constraints as targeted problem. Finally, Section \ref{sec:conclusion} concludes the whole study.  

\section{Background and Problem Description}
\label{sec:background}
\begin{figure}
\centering
\includegraphics[scale=0.45]{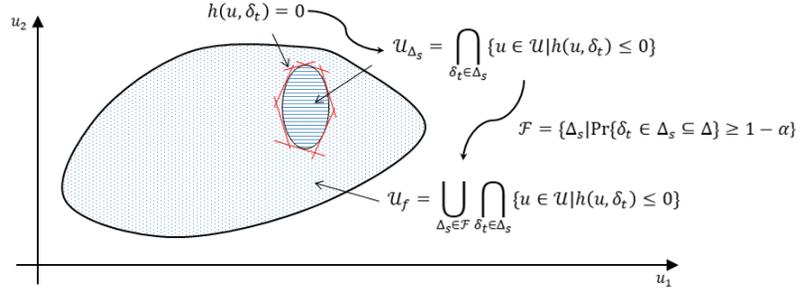}
\centering
\caption{Formulation of probabilistic feasible domain.}
\label{prob_fea_dom}
\end{figure}

Chance constrained program can be generally expressed as:
\begin{equation}
\label{eq_ccp}
\begin{split}
&\underset{u\in\mathscr{U}}{\text{min}} \,\, J(u) \\
&s.t.\quad  \text{Pr}\{h(u,\delta)\leq 0\}\geq 1-\alpha,\ \delta\in\Delta,\ \alpha\in (0,1) 
\end{split}
\end{equation} 
where $u\in\mathscr{U}\subset\mathbb{R}^{n_u}$ is the decision variable, the decision variable domain $\mathscr{U}$ is supposed to be bounded, $\delta\in\Delta\subset\mathbb{R}^{n_{\delta}}$ is an uncertain parameter vector, the set $\Delta$ is the sample space of $\delta$ on which a probability measure $\text{Pr}$ is defined, $\alpha$ is a given probability level for violation of chance constraints. Moreover, $J(u):\mathscr{U}\times\Delta\ \rightarrow\ \mathbb{R}$ and $\forall\delta\in\Delta,h(u,\delta):\mathscr{U}\times\Delta\ \rightarrow\ \mathbb{R}$ are continuous and differentiable in $u$. Chance constraints appearing in Eq. \eqref{eq_ccp} emerge in various applications and can be regarded as a compromise of hard constraints which require to be satisfied for all values $\delta\in\Delta$. Exploring optimizer under hard constraints could be too costly and even impossible. 

The feasible decision domain can be denoted as $\mathscr{U}_f$. However, it is difficult to obtain the exact expression of $\mathscr{U}_f$. Fig. \ref{prob_fea_dom} gives illustration on the formulation of probabilistic feasible domain in a 2-dimension space. Denoting $\Delta_{s}\subset\Delta$ and the probability measure of $\Delta_{s}$ satisfies 
\begin{equation}
\label{eq_pmds}
\text{Pr}\{\Delta_{s}\}\geq 1-\alpha.
\end{equation} 
The feasible domain for $\delta_t\in\Delta_{s}$ is defined as
\begin{equation}
\label{eq_fds}
\mathscr{U}_{\delta_t}=\{u\in\mathscr{U}|h(u,\delta_t)\leq 0\}.
\end{equation}
Then, the feasible domain for $\Delta_{s}$ is intersection of $\mathscr{U}_{\delta_t}$ for all $\delta_t\in\Delta_{s}$, which is written as  
\begin{equation}
\label{eq_fds}
\mathscr{U}_{\Delta_s}=\bigcap_{\delta_t\in\Delta_{s}}\mathscr{U}_{\delta_t}.
\end{equation}
Considering a family of $\Delta_s$ denoted as
\begin{equation}
\label{eq_family}
\mathscr{F}=\{\Delta_{s}\subset\Delta|\text{Pr}\{\Delta_{s}\}\geq 1-\alpha\},
\end{equation}
the feasible decision domain for Eq. \eqref{eq_ccp} can be defined as:
\begin{equation}
\label{eq_family}
\mathscr{U}_{f}=\bigcup_{\Delta_{s}\in\mathscr{F}}\mathscr{U}_{\Delta_s}=\bigcup_{\Delta_{s}\in\mathscr{F}}\bigcap_{\delta_t\in\Delta_{s}}\mathscr{U}_{\delta_t}.
\end{equation}
Obviously, even if $\mathscr{U}_{\delta_t}$ is known, it is impossible to obtain explicit expression or domain of $\mathscr{U}_{f}$ due to infinite times' operation of intersection and union. Thus, program \eqref{eq_ccp} is NP hard due to the chance constraint. To address program \eqref{eq_ccp}, the following issues should be considered:
\begin{itemize}
\item How to approximate chance constraints, namely approximate the probabilistic feasible domain of decision variable;
\item How to approximate the optimizer in the probabilistic feasible domain. 
\end{itemize}

\section{Preliminaries}
\label{sec:Preliminaries}
In this section, the preliminaries for understanding the proposed method is introduced. Firstly, a brief introduction on ELM algorithm is presented. Then, randomized optimization is introduced.

\subsection{Extreme Learning Machine}
ELM is essentially an algorithm to train the parameters in SLFNs. SLFNs is with a number of hidden nodes and with almost any nonlinear activation function, as an approximation of standard multilayer feedforward neural networks. For $N$ arbitrary distinct samples $(x_i,y_i)$, where $x_i=[x_{i,1},...,x_{i,m}]^T\in\mathbb{R}^n$ denotes the plant input and $y_i=[y_{i,1},...,y_{i,m}]^T\in\mathbb{R}^m$ the plant output, standard SLFNs with $\bar{N}$ hidden nodes and activation function $g(x)$ models the input-to-output relationship as
\begin{equation}
\label{eq_slfn}
\hat{t}_i=\sum_{j=1}^{\bar{N}}\beta_j g_j(x_i)=\sum_{j=1}^{\bar{N}}\beta_j g(\omega_j^T x_i + b_j),
\end{equation}
where $i=1,...,N$ is the sample index, $\omega_j=[\omega_{j,1},...,\omega_{j,n}]^T$ represent the weight vector connecting the $j$-th hidden node and the input nodes, $\beta_j=[\beta_{j,1},...,\beta_{j,m}]^T$ denotes the weight vector connecting the $j$-th hidden node and the output nodes, and $b_j$ is the threshold of the $j$-th hidden node. $\omega_{j}^Tx_{i}$ denotes the inner product of $\omega_{j}$ and $x_{i}$. The output nodes are linear in this SLFNs. 

The standard SLFNs with $\bar{N}$ hidden nodes with activation function $g(x)$ is able to approximate these $N$ samples with zero error means express as 
\begin{equation}
\label{eq_error}
\sum_{i=1}^{N}\left\lVert e_i \right\rVert=\sum_{i=1}^{N}\left\lVert t_i - \hat{t}_i \right\rVert=0, 
\end{equation}
i.e., according to the result in \cite{Huang2003}, there exit $\bar{N}\leq N,\beta_j, \omega_j$ and $b_j$ such that
\begin{equation}
\label{eq_slfn2}
t_i=\sum_{j=1}^{\bar{N}}\beta_j g(\omega_j^T x_i + b_j),
\end{equation}
where $j=1,...,N$ is sample index. More generally, by defining 
\begin{equation}
\label{eq_H}
H = 
\begin{bmatrix}
g(\omega_1^Tx_1+b_1) & ... & g(\omega_{\bar{N}}^Tx_{1}+b_{\bar{N}}) \\
\vdots & ... & \vdots \\
g(\omega_{\bar{N}}^Tx_{N}+b_1) & ... & g(\omega_{\bar{N}}^Tx_{N}+b_{\bar{N}}) \\
\end{bmatrix},
\end{equation}
\begin{equation}
\label{beta}
\beta=
\begin{bmatrix}
\beta_1^T \\
\vdots \\
\beta_{\bar{N}}^T \\
\end{bmatrix},
\end{equation}
and
\begin{equation}
\label{T}
T=
\begin{bmatrix}
t_1^T \\
\vdots \\
t_{\bar{N}}^T \\
\end{bmatrix},
\end{equation}
 the above $N$ equations can be written compactly as
\begin{equation}
\label{eq_slfn3}
T=H\beta.
\end{equation}
As introduced in \cite{Huang1998}, $H$ is called the hidden layer output matrix of the neural network; the $i$-th column of $H$ is the $i$-th hidden node output with respect to inputs $x_1, x_2,...,x_N$. The gradient-based algorithm can be used to train the value of $\beta, \omega_1,...,\omega_{\bar{N}},b_1,...,b_{\bar{N}}$ \cite{Tamura}. Compare to the gradient-based algorithm, ELM algorithm, proposed in \cite{Huang2006}, is more simple and efficient. 
Regarding $X=[x_1,x_2,...,x_N]$ as input, and $T$ as output, the Batch ELM algorithm for training the SLFNs models can be summarized as in algorithm 1:  
\begin{algorithm}[h]
\label{alg1}
\caption{Batch ELM for training the SLFNs models} 
\textbf{Input:} $X,T,\bar{N}$   \\
\textbf{Output:} $\beta,\omega_{j}, b_{j},j=1,2,...,\bar{N}$ \\
\ \ 1:\ Step 1: Randomly assign $\omega_{j}$ and $b_{j}$, $j=1,...,\bar{N}$ \\
\ \ 2:\ Step 2: Calculate the hidden layer output matrix $H$ \\
\ \ 3:\ Step 3: Calculate the $\beta$ as $\beta=H^{M}T$. 

\end{algorithm}

In algorithm 1, $H^{M}$ is the Moore-Penrose generalized inverse of matrix $H$ \cite{Rao}, which can be derived by
\begin{equation}
\label{eq_HM}
H^{M}=(H^{T}H)^{-1}H^{T}.
\end{equation}
Then, the estimation of $\beta$ can be calculated as 
\begin{equation}
\label{eq_HM}
\beta=(H^{T}H)^{-1}H^{T}T.
\end{equation}

 The basis of algorithm 1 is the results presented in \cite{Tamura}, if the activation function $g$ is infinitely differentiable the hidden layer output matrix $H$ is invertibel and $\left\lVert H\beta-T \right\rVert=0$. The sequential implementation of Eq. \eqref{eq_HM} can be derived and referred as the recursive least squares (RLS) algorithm. The proof of the RLS algorithm can be found in \cite{Chong}. The sequential ELM for training the SLFNs models is derived based on RLS algorithm and summarized as in algorithm 2:
\begin{algorithm}[h]
\label{alg2}
\caption{Sequential ELM for training the SLFNs models} 
1:\ Step 1: Give $\beta_0$ according to algorithm 1  \\
2:\ Step 2: Calculate the hidden layer output matrix $h_{k+1}$ based on further coming observation $(x_{k+1},t_{k+1})$ according to Eq.\eqref{eq_H}, $k=0,1,2,...,i,...$ \\
3:\ Step 3: Calculate the $\beta_{k+1}$ as
\begin{equation}
\label{eq_SELM}
\beta_{k+1}=\beta_{k}+M_{k+1}h_{k+1}(t_{k+1}^{T}-h_{k+1}^T\beta_{k})
\end{equation} 
where $M_{k+1}$ is calculated as
\begin{equation}
\label{eq_M}
M_{k+1}=M_{k}-\frac{M_kh_{k+1}h_{k+1}^TM_k}{1+h_{k+1}^TM_kh_{k+1}}.
\end{equation}
\\
4:\ Step 4: Set $k=k+1$
\end{algorithm}

\subsection{Randomized Optimization Algorithm}

Randomized optimization algorithm is numerical optimization algorithm which is free from the gradient of the objective function\cite{Matyas}. It can therefore be used for problems with objective functions that are not continuous or differentiable. Furthermore, even if the problem is under non-convex constraints, randomized optimization algorithm still ensures the convergence to the optimizer \cite{Baba}. In random optimization, sample of decision is extracted from the feasible domain, the start point can be randomly chosen. Then, the next sample is extracted near the current position according to normal distribution or uniform distribution. The optimizer will be replaced if better sample which has lower cost or better reward is extracted. It is ensured that the sample moves to better positions iteratively and stops at global optimizer\cite{Dorea}.

Let $F:\mathscr{V}\subset\mathbb{R}^n\rightarrow\mathbb{R}$ be the cost function to be minimized in the feasible domain $\mathscr{V}$. Let $v\in\mathscr{V}$ denote a position or candidate optimizer in the feasible domain. The basic randomized optimization algorithm is denoted as algorithm 3 and expressed as following\cite{Solis}:
\begin{algorithm}[h]
\label{alg3}
\caption{Typical randomized optimization algorithm} 
1:\ Step 1: Initialize $v\in\mathscr{V}$ randomly  \\
2:\ Step 2: Extract new sample $v'\in\mathscr{V}_v$ obeying uniform distribution or normal distribution. Here, $\mathscr{V}_v$ is a neighbour of $v$, for instance, as $
\mathscr{V}_v=\{v_f\in\mathscr{V}|\left\|v_f-v\right\|^2<\epsilon_v\}$ \\
3:\ Step 3: If $F(v')<F(v)$, set $v=v'$ \\
4:\ Step 4: Examine whether the termination criterion is met(e.g. number of iterations), if termination criterion is met, go to Step 5, otherwise, go to Step 2 and repeat \\
5:\ Step 5: Set $v$ as optimizer
\end{algorithm}


\section{Proposed Method}
\label{sec:proposed}
This section discusses the two-phase optimizer exploration approach for chance constrained program. Firstly, the structure of the proposed method is presented. Then, approximate algorithm for violate probability is introduced. Finally, the exploration algorithm is introduced in details.

\subsection{Overview}

\begin{figure}
\centering
\includegraphics[scale=0.45]{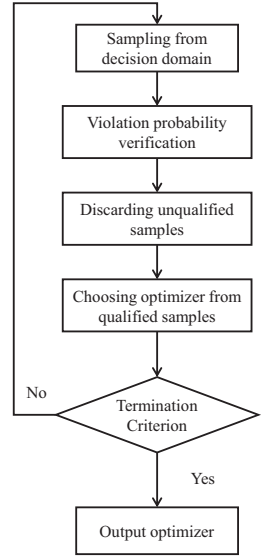}
\centering
\caption{Flow diagram of randomized optimization for chance constrained program.}
\label{ran_fra}
\end{figure}

Randomized optimization can be used to solve program \eqref{eq_ccp}. The flow diagram of randomized optimization applied for chance constrained program is illustrated in Fig. \ref{ran_fra}. Firstly, decision samples are randomly extracted from $\mathscr{U}$ which might not satisfy the chance constraints. Thus, in the next step, violation probability of extracted samples are verified and the unqualified samples are discarded. The optimizer are chosen from the qualified samples. The process is repeated until termination criterion is satisfied and optimizer is outputted finally. The first key of the optimization algorithm is to have a map from decision to violation probability. The ELM-based approximate approach for violation probability is proposed in this study and introduced in the next part. The details of optimizer exploration algorithm is also presented in the later part of this section.

\subsection{Violation Probability Map Approximation}

\begin{figure}
\centering
\includegraphics[scale=0.45]{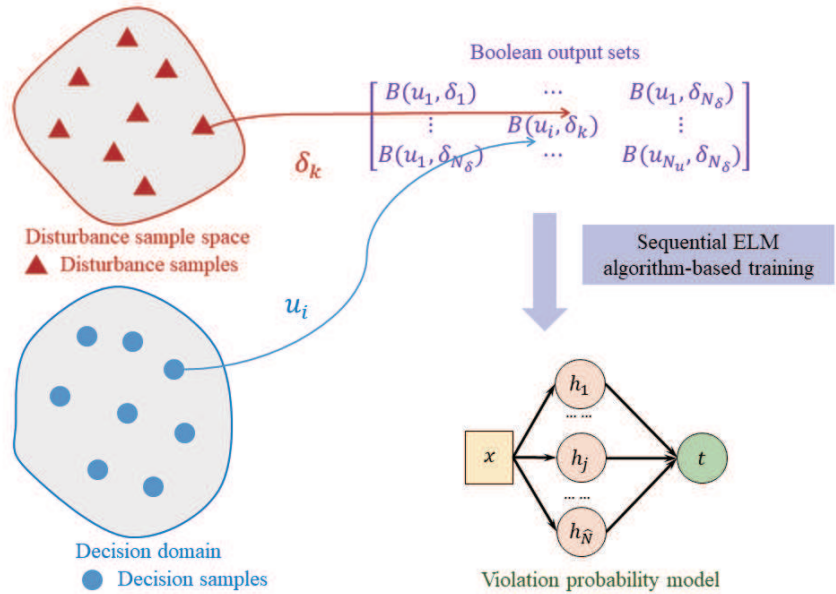}
\centering
\caption{Brief structure of approximate algorithm.}
\label{app_alg}
\end{figure}

Violation probability of decision $u$ is denoted as 
\begin{equation}
\label{eq_vioprob}
V(u)=\text{Pr}\{h(u,\delta)>0\},\delta\in\Delta.
\end{equation}
Obviously, $V(u)$ is a function of $u$ since the fixed $u$ have only one violation probability value when probability measure of $\Delta$ is identical. Then, if $V(u)$ is continuous, it can be approximated by SLFNs model. 

For given decision $u$ and disturbance $\delta$, a boolean function can be denoted as 
\begin{equation}
\label{eq_bool}
B(u_c,\delta)=
\left\{
\begin{split} 
&1, \text{if}\ h(u,\delta)>0 \\
&0, \text{if}\ h(u,\delta)\leq0 
\end{split}
\right.
\end{equation}
If $N_s$ samples, $\{\delta_{1},...,\delta_{N_s}\}$, are extracted from $\Delta$ independently, the estimate violation probability can be written as 
\begin{equation}
\label{eq_vpest}
\hat{V}(u) = \frac{\sum_{i=1}^{N_s}B(u,\delta_i)}{N_s}.
\end{equation}
According to law of larger numbers, as $N_s\rightarrow\infty$, $\hat{V}(u)$ converges to $V(u)$ with probability one\cite{Feller}. 

For $N_u$ samples in decision domain, denoted as $u_{1},...,u_{N_u}\in\mathscr{U}$, can be regarded as input of a SLFNs model. The corresponding violation probability values, $V(u_{1}),...,V(u_{N_u})$ can be regarded as output of the SLFNs model. Although the real values of violation probability are not known, it can be iteratively approximated by Eq. \eqref{eq_bool}. Thus, for every decision sample $u_{i},i\in\{1,...,N_u\}$, it has an boolean output $B(u_{i},\delta_{k}),k\in\{1,...,N_{\delta}\}$ for determined disturbance sample which can also be used as the output when updating the violation probability map. The approximate algorithm is summarized in Algorithm 4, which is essentially a two-layer sample algorithm. Also, brief structure of approximate algorithm is illustrated in Fig. \ref{app_alg}. If $N_u,N_{\delta}\rightarrow\infty$, the algorithm converges to the violation probability map.
\begin{algorithm}[h]
\label{alg4}
\caption{Algorithm for approximating violation probability map} 
1:\ Step 1: Initialize parameters, $\beta,\omega_j,b_j,j=1,...,\hat{N}$ randomly\\
2:\ Step 2: Extract decision samples $u_1,...u_{N_u}\in\mathscr{U}$ obeying uniform distribution, the decision samples are fixed from the following steps \\
3:\ Step 3: Extract sample $\delta_{k+1}\in\Delta$ obeying the distribution defined by probability measure $\text{Pr}\{\cdot\}$\\
4:\ Step 4: Calculate $B(u_{i},\delta_{k+1}),\forall i\in\{1,...,N_u\}$ according to Eq. \eqref{eq_bool}\\
5:\ Step 5: Regard $u_{i},B(u_{i},\delta_{k+1})$ as $x_{k+1},t_{k+1}$ in algorithm 2, then, update $\beta$ for violation probability map according to algorithm 2 for all $i\in\{1,...,N_u\}$ \\
6:\ Step 6: Set $k=k+1$
\end{algorithm}

\subsection{Optimizer Exploration Algorithm}
\begin{figure}
\centering
\includegraphics[scale=0.45]{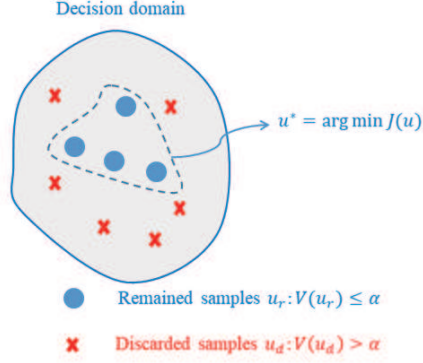}
\centering
\caption{Brief structure of optimizer exploring algorithm.}
\label{exp_opt}
\end{figure}
The proposed optimizer exploration algorithm firstly extracts $n$ samples, $\{u_{1},...,u_{n}\}\in\mathscr{U}$ , from the decision domain. However, not all the samples satisfy the chance constraints. Based on the trained violation probability model, $V(u_{i})=Pr\{h(u_{i},\delta)>0\}, \forall i\in\{1,...,n\}$ can be calculated and the ones with violation probability larger than $\alpha-\alpha_{\epsilon}$ are discarded. $\alpha_{\epsilon}\in(0,\alpha)$ is chosen for technical consideration since the violation probability map may have errors due to that $N_{\delta},N_{u}$ cannot be $\infty$. Thus, more conservative boundary than the required one should be chosen. The remained feasible samples are used to calculate the corresponding cost function values. The one with minimal cost value is chosen to be the optimizer. The above steps are repeated until termination criterion is met, for instance, number of iterations. The algorithm is summarized in Algorithm 5 and  briefly illustrated in Fig. \ref{exp_opt}. 
\begin{algorithm}[h]
\label{alg5}
\caption{Algorithm for exploring the feasible optimizer} 
1:\ Step 1: Initialize optimizer $u_f^{*}\in\mathscr{U}$ randomly\\ 
2:\ Step 2: Extract $\{u_{1},...,u_{n}\}\in\mathscr{U}$ randomly\\
3:\ Step 3: Calculate $Pr\{h(u_{i},\delta)>0\}, \forall i\in\{1,...,n\}$, discard all $u_{i}$ that $V(u_{i})=Pr\{h(u_{i},\delta)>0\}>\alpha$, and the remained solution set is $\mathbb{U}_f$ \\
4:\ Step 4: If $\mathbb{U}_f=\emptyset$, return to Step 1, else if $\mathbb{U}_f\neq\emptyset$, go to Step 5\\
5:\ Step 5: Set optimizer as $u^{*}=\underset{u\in\mathbb{U}_f}{\text{argmin}} J(u)$\\
6:\ Step 6: Replace optimizer $u_f^{*}=u^{*}$ and go to Step 7 if $J(u^{*})<J(u_f^{*})$; Return to Step 2 directly if $J(u^{*})\geq J(u_f^{*})$  \\
7:\ Step 7: Examine whether the termination criterion is met (e.g. number of iterations), if termination criterion is met, go to Step 8, otherwise, go to Step 2 and repeat \\
8:\ Step 8: Set $u_f^{*}$ as optimizer.
\end{algorithm}

\section{Numerical Validation}
\label{sec:numerical simulation}

\begin{figure}
\centering
\includegraphics[scale=0.45]{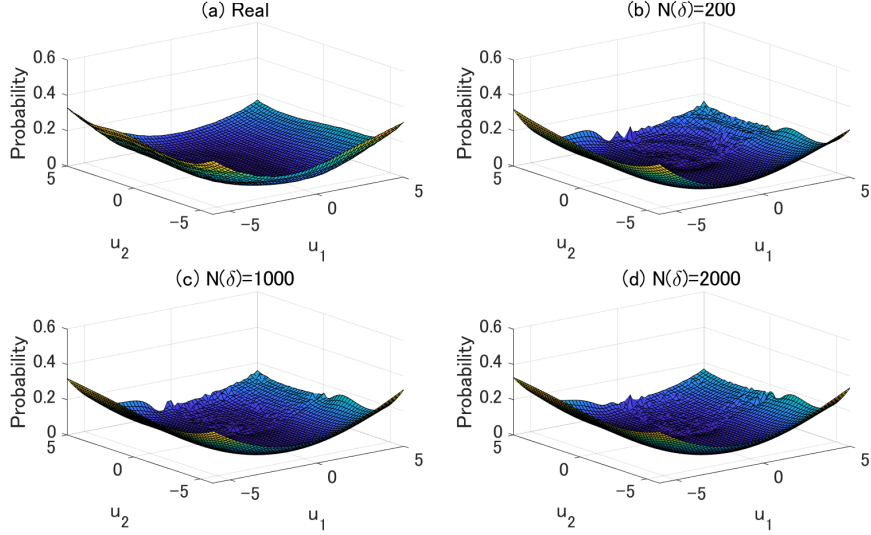}
\centering
\caption{Approximation of violation probability map.}
\label{feasible_domain}
\end{figure}

\begin{figure}
\centering
\includegraphics[scale=0.45]{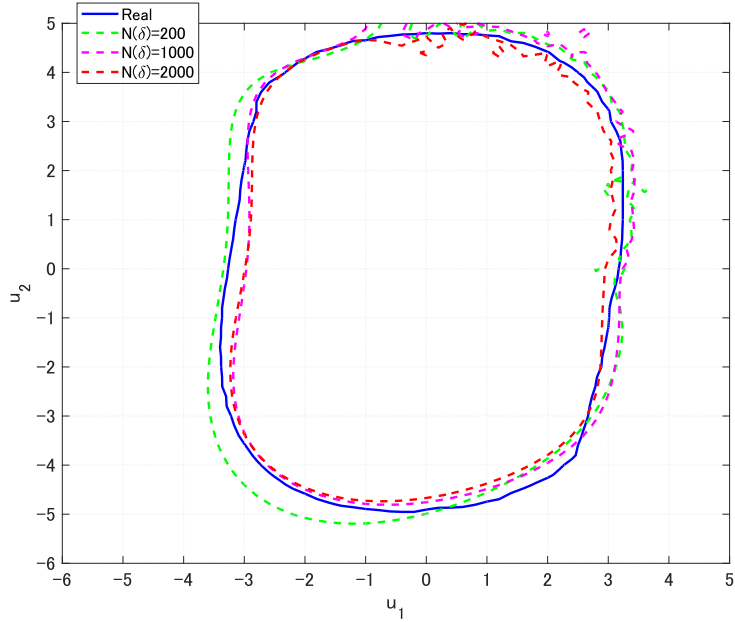}
\centering
\caption{Boundaries of $5\%$ violation with different $\delta$-sample numbers.}
\label{boundary}
\end{figure}

This section presents numerical simulation for verifying the proposed method, comparing with scenario approach and parallel randomized algorithm. The targeted problem is briefly introduced firstly. Also, the basic concepts of scenario approach and parallel randomized algorithm are introduced shortly which is enough for helping understand the comparison result. Furthermore, the results of both probabilistic feasible domain approximation and optimizer exploration are presented. 

\subsection{Targeted Problem}

The targeted problem in the numerical simulation is a non-convex program with chance constraints. The decision domain is $\mathscr{U}=[-6,5]^2$. The cost function is 
\begin{equation}
\label{eq_tp_cost}
J(u) = \frac{\sum_{i=1}^2(u_i+0.5)^4-30u_i^2-20u_i}{100}.
\end{equation}
The constrained function is
\begin{equation}
\label{eq_tp_cons}
h(u,\delta) = (\sum_{i=1}^2 0.075*(u_i-2\delta)^4-3.5*(u_i-2\delta)^2)-(8-0.1\delta)^2,
\end{equation}
where $\delta$ is random variable which obeys normal distribution $\mathscr{N}(0,1)$. Moreover, the violation probability level is $\alpha=0.05$. The proposed method is applied to solve the targeted problem, comparing with scenario approach presented in \cite{CampiSampling} and parallel algorithm proposed in \cite{Shen2019}. 

In scenario approach, independent samples $\delta^{(i)},i=1,...,N$ is identically extracted from $\Delta$ randomly, a deterministic convex optimization problem can be formed as
\begin{equation}
\label{eq_dcop}
\begin{split}
&\underset{u\in\mathscr{U}}{\text{min}} \,\, J(u) \\
&s.t.\quad  h(u,\delta^{(i)})\leq 0,\ i=1,...,N
\end{split}
\end{equation}
which is a standard finitely constrained optimization problem. The optimal solution $\hat{u}_N$ of the program Eq. \eqref{eq_dcop} is called as the scenario solution for program Eq. \eqref{eq_ccp} generally. Moreover, since the extractions $\delta^{(i)},i=1,...,N$ is randomly chosen, the optimal solution $\hat{u}_N$ is random variable. If $\hat{u}_N$ is expected to satisfy
\begin{equation}
\label{eq_sa_pr}
\text{Pr}^{N}\{(\delta^{(1)},...,\delta^{(N)}\in\Delta^N:V(\hat{u}_N)\leq\alpha\}\geq1-\beta, \beta\in(0,1),
\end{equation}
then, $N$ should have a lower limitation $N_l$
\begin{equation}
\label{eq_sa_n}
N\geq \frac{2}{\alpha}\text{ln}\frac{1}{\beta}+2n_u+\frac{2n_u}{\alpha}\text{ln}\frac{2}{\alpha}.
\end{equation}
Note that $\beta$ is an important factor and choosing $\beta=0$ makes $N_l=\infty$. Namely, if the number of chosen samples gets larger, the probability of satisfying the original chance constraints approaches 1. Actually, when chosen samples becomes infinity and cover the whole sample space, not only the chance constraints, it becomes totally robust.

Parallel randomized algorithm is under the same framework of the proposed method. Differently, the discard process of parallel randomized algorithm also adopts sampling approach. Parallel randomized algorithm is summarized in algorithm 6.
\begin{algorithm}[h]
\label{alg6}
\caption{Paralle randomized algorithm} 
1:\ Step 1: Initialize optimizer $u_f^{*}\in\mathscr{U}$ randomly\\ 
2:\ Step 1: Extract $\{u_{1},,...,u_{i},...,u_{N_u}\}\in\mathscr{U}$ \\
3:\ Step 2: Extract $\{\delta_{1},...,\delta_{N_{\delta}}\}\in\Delta$ randomly\\
4:\ Step 3: Calculate approximately $V(u_{i}), \forall i\in\{1,...,N_u\}$ according to Eq. \eqref{eq_bool} and \eqref{eq_vpest}, discard all $u_{d}$ that $V(u_{d})>\alpha$, and the remained solution set is $\mathbb{U}_f$ \\
5:\ Step 5: If $\mathbb{U}_f=\emptyset$, return to Step 1, else if $\mathbb{U}_f\neq\emptyset$, go to Step 6\\
6:\ Step 6: Set optimizer as $u^{*}=\underset{u\in\mathbb{U}_f}{\text{argmin}} J(u)$\\
7:\ Step 7: Replace optimizer $u_f^{*}=u^{*}$ and go to Step 8 if $J(u^{*})<J(u_f^{*})$; Return to Step 2 directly if $J(u^{*})\geq J(u_f^{*})$  \\
8:\ Step 8: Examine whether the termination criterion is met (e.g. number of iterations), if termination criterion is met, go to Step 9, otherwise, go to Step 1 and repeat \\
9:\ Step 9: Set $u^{*}$ as optimizer.
\end{algorithm} 

\subsection{Probabilistic Feasible Domain}

Firstly, the results of probabilistic feasible domain estimation by proposed method is presented. The violation probability maps, comparison between real one and approximations, are shown in Fig. \ref{feasible_domain}. Actually, the real one is also the approximate one. 3136 points, $(u_{1,i},u_{1,i})$ are chosen from $[-6,5]^2$. For every point, 10000 samples of $\delta$ are extracted according to $\mathscr{N}(0,1)$ for calculating the corresponding violation probability. Fig. \ref{feasible_domain}(a) shows the results. Fig. \ref{feasible_domain} (b-d) shows the results of approximation maps based on Algorithm 4 with $N(\delta)=200$, $N(\delta)=1000$ and $N(\delta)=2000$ where $N(\delta)$ denotes the sample numbers. Since the boundary of $5\%$ violation probability is concerned, the comparisons of boundaries are plotted in Fig. \ref{boundary}. Obviously, Algorithm 4 converges to the real boundary when $N(\delta)\geq 1000$.

\subsection{Optimizer Exploration}
\begin{figure}
\centering
\includegraphics[scale=0.45]{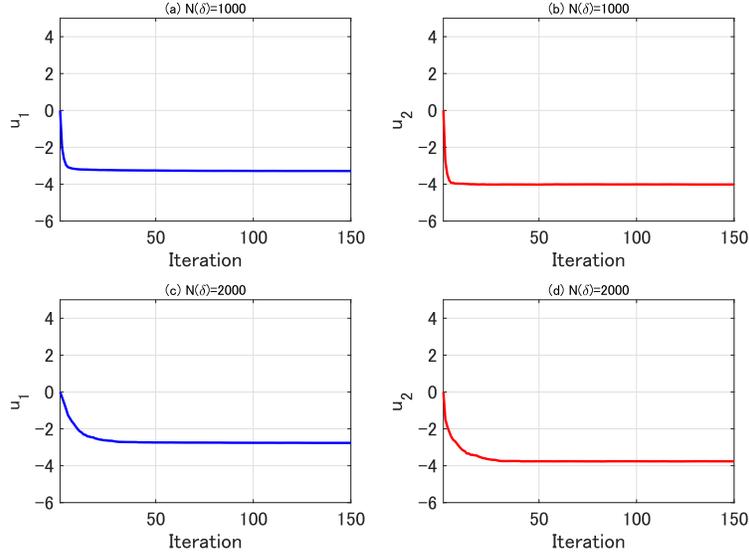}
\centering
\caption{Mean evolutions of the solution during computation process for proposed method (simulation of 400 times).}
\label{u_evo}
\end{figure}

\begin{figure}
\centering
\includegraphics[scale=0.45]{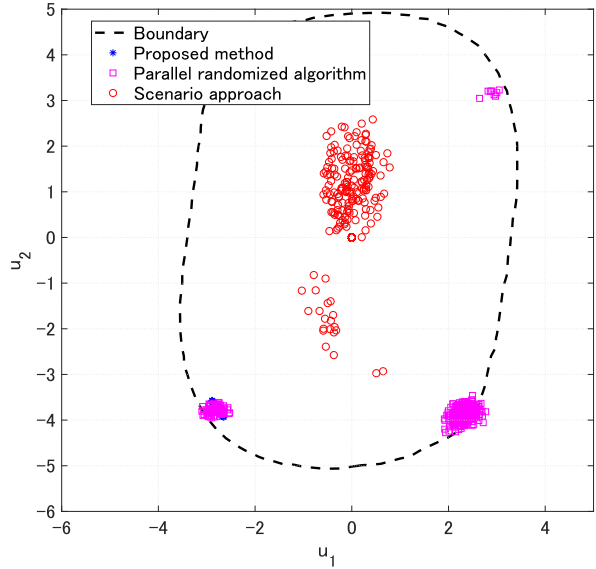}
\centering
\caption{Solution distribution ($N_{\delta}=2000$,simulation of 400 times).}
\label{Sol_dis}
\end{figure}

\begin{figure}
\centering
\includegraphics[scale=0.45]{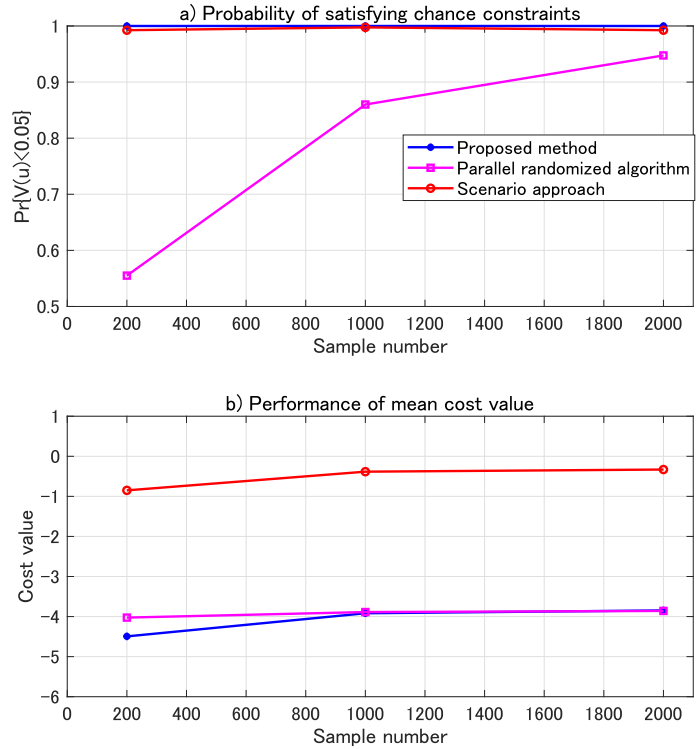}
\centering
\caption{Results of the cost values and violation probabilities (mean values of simulation of 400 times).}
\label{res_plo}
\end{figure}

The results of optimizer exploration are presented here. In this numerical simulation, proposed method, parallel randomized algorithm and scenario approach are used to solve the targeted problem. The numerical simulation adopts Monte Carlo methodology, namely, the procedure of exploring the optimizers is repeated for a large number of times and the samples of $u$ and $\delta$ are extracted obeying the identical distribution in every simulation time. 400 times of simulation were done in this validation. Moreover, $\alpha_{\epsilon}$ is chosen as 0.005 here.

In Fig. \ref{u_evo}, mean evolutions of the solution during computation process for proposed method are plotted, for both cases $N(\delta)=1000$ and $N(\delta)=2000$. In every simulation, the solutions converge to the small intervals for both $u_1$ and $u_2$. The final converged values of decision are plotted in Fig. \ref{Sol_dis}. Blue points are solutions from proposed method using 2000 disturbance samples, magenta squares are solutions from parallel randomized algorithm using 2000 disturbance samples and red circles present the solutions from scenario approach using 2000 disturbance samples. Apparently, proposed method can converges to the $5\%$ boundary where the cost is minimal in the probabilistic domain. Parallel randomized algorithm exhibits worse performance than proposed method while roughly near the boundary with randomness. However, for scenario approach, the solution cannot converge to the boundary. In every simulation, the sampled $\delta$ are different, the constraints of problem expressed by Eq. \eqref{eq_dcop} are different, the solution are therefore different and distributed in a larger area. 

Furthermore, Fig. \ref{res_plo} shows the plotted mean results of cost values and violation probabilities in all simulations. From these results, obviously, the proposed method achieves the trade-off between violation probability and optimization in a more stable way because SLFN-based violation probability map is accurate. While, scenario approach achieves the chance constraints but get poor performance on cost value since it is too conservative. Parallel randomized algorithm achieves good performance on cost value but has poor performance on satisfying violation probability since its estimation on violation probability has variance and have poor accuracy than the SLFN-based map. 

\section{Conclusion}
\label{sec:conclusion}

This paper has introduced a neural network-based approximate approach to chance constrained optimization. The novel idea is to approximate the feasible domain constrained by chance constraints by SLFNs. The parameters of SLFNs can be obtained by ELM algorithm with historical data. Then, a two-layer randomized optimization algorithm is proposed to approximate the optimizer of chance constrained program based on the trained probabilistic feasible domain map. The proposed method is validated by numerical simulation compared with scenario approach and parallel randomized algorithm. The proposed method exhibits better robustness on both exploring optimizer and satisfying violation probability. while, there still remains future works to be done for improving the proposed method. The current randomized optimization algorithm is totally a random one which can be improved to converge to the optimizer in less iterations. Moreover, the accuracy of probability map is related to the sample number of random disturbance. The quantitative analysis should be implemented to investigate the relationship between accuracy and sample number. Especially, the principle of choosing parameter $\alpha_{\epsilon}$ should be investigated in theory. 

\bibliographystyle{unsrt}  


\end{document}